\documentstyle[epsfig]{mn}
\newcommand{\lfir}{L_{\rm FIR}}
\newcommand{\lir}{L_{\rm IR}}
\newcommand{\lb}{L_{\rm B}}
\newcommand{\lco}{L_{\rm CO}}
\newcommand{\msol}{M_{\sun}}
\newcommand{\lsol}{L_{\sun}}
\newcommand{\mbetabest}{M_{850\mu{\rm m}}}
\newcommand{\mco}{M_{\rm CO}}
\newcommand{\apj}{ApJ, }
\newcommand{\aajou}{A\&A, }
\newcommand{\mn}{MNRAS, }

\begin{document}

\title[Dust and gas in luminous infrared galaxies]
{Dust and gas in luminous infrared galaxies - results from 
SCUBA observations}
\author[U. Lisenfeld, K.G.Isaak \& R. Hills]
{U. Lisenfeld$^1$, K.G. Isaak$^{2}$\thanks{Now at 
Cavendish Astrophysics Group, Cavendish Laboratory, Madingley Road, 
Cambridge CB3 OHE, UK, e-mail: isaak@mrao.cam.ac.uk}
, R. Hills$^3$ \\ 
$^1$IRAM, Avenida Divina Pastora 7, N.C., 
18012 Granada, Spain, e-mail: ute@iram.es\\
$^2$Department of Astronomy, University of Maryland, College Park, USA \\
$^3$Cavendish Astrophysics Group, Cavendish Laboratory, 
Madingley Road, Cambridge CB3 OHE, UK }

\maketitle
\begin{abstract}
We present new data taken at 850 $\mu$m with SCUBA at the JCMT  for a sample
of 19 luminous infrared galaxies. Fourteen galaxies were detected. 
We have used these data, together with fluxes at 25, 60 and 
100 $\mu$m from IRAS, to model the dust emission. We find that 
the emission from most galaxies can be described by an optically thin, 
single temperature dust model with an exponent of the dust extinction 
coefficient ($k_\lambda \propto \lambda^{-\beta}$) of $\beta \simeq 1.4 - 2$.
A lower $\beta\simeq 1$ is required to model the dust emission
from two of the galaxies, Arp 220 and NGC 4418. We discuss
various possibilities for this difference and conclude that the most
likely is a high dust opacity. In addition, we compare
the molecular gas mass derived from the dust emission, $\mbetabest$, 
with the molecular gas mass derived from the CO emission, $\mco$, and 
find that $\mco$ is on average a factor 2--3 higher than $\mbetabest$.
\end{abstract}


\section{Introduction}
Much observational evidence has been accumulated since the first discovery of
luminous (infrared luminosities of $\lir=L(8-1000\mu{\rm m}) >10^{11} 
\lsol$, ``LIRGs'')  and ultraluminous ($\lir >10^{12}  
\lsol$, ``ULIRGs'') infrared galaxies to suggest that their interstellar 
medium (ISM) is quite different from that of normal galaxies.
ULIRGs and LIRGs contain very large quantities of molecular gas, 
as traced by CO (e.g. Sanders et al. 1988), and by tracers of 
high density gas such as HCN and CS(3-2) (Solomon, Downes \& 
Radford 1992).
In spite of their abundance of molecular hydrogen, LIRGs have ratios 
of far-infrared (FIR) luminosity to molecular gas mass, $\lfir/M_{\rm H_2}$, 
that largely exceed those measured in our own galaxy \cite{solomon92}.  
The ratio between the luminosity of the high-density tracers
and $\lfir$, however, is quite normal. The FIR emission is a measure
of the star formation activity in a galaxy and so the constant 
$\lfir/L_{\rm HCN}$  implies that the star formation rate (SFR) per 
{\it dense} gas mass is normal. The ratio of the luminosity 
of the high-density gas tracers to CO is much higher than in our own galaxy, 
and so it would appear that a larger fraction of the molecular gas 
is in a dense state. Solomon et al. \shortcite{solomon97} 
conclude from this that the 
ISM of ULIRGs is much denser than in normal galaxies, with even 
gas in intercloud regions being in the molecular state.

The dust properties of LIRGS are also different from those found in less
luminous galaxies.
There are a number of strong indications that
the dust opacity in LIRGs is very high, with the result that the emission from
LIRGs is optically thick even in the FIR wavelength range. 
Based on the 
tight correlation between the flux at 100 $\mu$m
and the CO emission of ULIRGs, Solomon et al. \shortcite{solomon97} 
argue that the ISM is 
optically thick at wavelengths as long as 100 $\mu$m. A high 
opacity in Arp 220 is supported by a range of observations: 
(i) Emerson et al. \shortcite{emerson}  have concluded from observations
in the IR and submillimetre
that the dust opacity is of order unity between 100 and 200 $\mu$m.
(ii) From observations with ISO using the 
Long Wavelength Spectrometer (LWS) Fischer et al. \shortcite{fisher} derived 
unity optical  depth at $\simeq 150 \mu$m. (iii) The ratios 
of the IR fine structure line emission observed by ISO \cite{genzel} 
are consistent with a screen opacity of $A_V \simeq 45$  mag, 
equivalent to an opacity of $A_V \simeq 1000$ mag if the dust and gas were 
well-mixed.  
(iv) Calculations by Downes \& Solomon \shortcite{downes} 
using measurements of 
the CO surface brightness and a galactic dust-to-gas ratio  suggest 
an opacity of about $A_V \simeq 1000$ mag. 
The dust opacity is also high in NGC 4418, where a high dust 
extinction has been measured at optical wavelengths ($A_V \gg 50$ mag, 
Roche et al. 1986). 
The high dust opacity might also explain why the  
C$^+$ 158 $\mu$m line is so weak in ULIRGs (Malhorta et al. 1997,
Luhman et al. 1998).

An accurate measurement of the molecular gas mass of a galaxy is critical 
to determining star formation activity. Estimating the molecular gas mass 
of an object is not straightforward, and is typically made in one of two ways: 
The first method relies on observations of the optically thick 
CO(1-0) emission. The calculated CO line luminosity can be 
converted to a $\rm H_2$ gas mass by adopting a conversion factor between 
CO and $\rm H_2$ that has been derived from observations of Galactic 
molecular clouds (e.g. Young \& Scoville 1982) and $\gamma$-rays fluxes
\cite{bloemen}. 
The second method makes use of the empirical relationship between gas and 
dust mass. The dust mass can be determined from the dust emission spectrum
with assumptions made about the dust emissivity and composition. The dust mass
can in turn be converted into a molecular gas mass using the locally 
derived gas-to-dust mass ratio  (e.g. Hildebrand 1983).

Both derivations assume that the dust and gas parameters derived from 
measurements in our own galaxy are valid under the quite different 
conditions in LIRGs, and that they are independent of star formation activity.
Theoretical arguments by Maloney \& Black \shortcite{maloney}
suggest that the 
CO-to-H$_2$ conversion factor should be considerably lower in actively 
star forming galaxies because of the higher density and 
temperature of the constituent molecular clouds.
Observationally, there are also indications that the Galactic 
CO-to-H$_2$ conversion factor overestimates the molecular gas mass 
in LIRGs: Shier, Rieke \& Rieke \shortcite{shier} find that the molecular
gas mass derived from CO emission using the standard Galactic conversion 
factor exceeds the dynamical mass by 
a factor of 4 -- 10. Similarly, Solomon et al. \shortcite{solomon97} 
conclude  from comparisons of the dynamical
and gas mass derived from the FIR emission and the CO,  
that the Galactic conversion factor between the CO luminosity,
$\lco$, and the molecular gas mass, $M_{\rm H_2}$, overestimates  
$M_{\rm H_2}$ by a factor of about 5. This conclusion is
supported by a better determination of the dynamical mass of a sample
of ULIRGs using high angular resolution observations made with 
the Plateau de Bure interferometer \cite{downes}. 

In this paper we report on observations of the 850 $\mu$m 
emission for a sample of 19 LIRGS galaxies using the 
submillimetre-wave continuum bolometer array, SCUBA \cite{holland}, 
at the JCMT\footnote
{The James Clerk Maxwell Telescope is operated by The Joint Astronomy 
Centre on behalf of the Particle Physics and Astronomy Research Council of 
the United Kingdom, the Netherlands Organization for Scientific Research, 
and the National Research Council of Canada.}.  
The aim of these observations has been twofold: firstly, to 
study the dust emission from a selection of galaxies believed to 
have an ISM quite different to that of our own Galaxy, and secondly, to 
compare the molecular gas mass derived from the dust emission at 
850 $\mu$m with that derived from CO measurements in order to assess 
whether the assumptions on which both
methods rely are valid for galaxies spanning a wide range of 
different star formation activities. 
The large field of view of SCUBA enables us to include any extended
dust emission and to measure its size. In this way we are able to avoid
the limitations of previous studies made with single-pixel bolometers
(e.g. Andreani \& Franceschini 1996, Lisenfeld et al. 1996).

\section{Sample and observations}

The sample consists of 19 galaxies selected from the subsample of the
Bright IRAS galaxies observed in CO(1-0) with the NRAO 12m
telescope by Sanders et al. \shortcite{sanders91}. 
The combination of the large CO beam (55'')
and the small optical diameter of our 
sample means that no CO emission has been missed.
The galaxies were selected to cover a wide range of infrared luminosities
($\rm 7.6\, 10^{10} \lsol < \lir < 3.1\, 10^{12} \lsol$)
\footnote{We follow the definition of
IR luminosity of Sanders \& Mirabel \shortcite{sanders96} with
$\lir=L(8-1000\mu{\rm m})=5.6\,10^5 D_{\rm Mpc}^2(13.48 F_{12}+ 5.16 F_{25}+
2.58 F_{60}+F_{100})$ where the fluxes are in Jy.} 
and 
infrared-to-blue-luminosity ratios $\lir/\lb$ ($2 < \lir/\lb < 80$), 
thus sampling a wide range of luminosities and star formation 
activities (as traced by $\lir/\lb$).

The galaxies were mapped in service time with SCUBA at the
JCMT between July 1997 and April 1998. 
The 16-jiggle mode was used for the observations to obtain
fully sampled maps at 850 $\mu$m; the weather conditions were
too poor in order to make use of the 450 $\mu$m data that was
obtained parallely. The chop throws ranged betweeen
60'' and 120'' (see Table 1) and the integration time per map
was between 15  and 70 minutes. The sky opacity at 850 $\mu$m varied
between 0.2 and 0.45.
The data were reduced using the software package SURF 
\cite{jenness}. The data were flatfielded, extinction-corrected and 
despiked, with bad bolometers and bad channels removed prior to 
rebinning and flux calibration. Flux calibration was achieved using Uranus 
and  the secondary 
calibrators CRL~618, IRC+10216, OH231.8. 
The conversion factor between volts and Jy was found to be quite consistent
over different nights, with an average value of $264\pm 19$ Jy/V.

\section{Results}

Listed in Table 1 are the measured 850 $\mu$m fluxes (column 6) together with 
12, 25, 60 and 
100 $\mu$m IRAS fluxes taken from the IRAS Faint Source Catalog 
\shortcite{IRAS}. 
The flux errors given in column 6 represent the noise in the
maps, $\sigma_{\rm n}$. The total error, $\sigma$, used for the model fitting,
takes additional account of the estimated calibration error, 
$\sigma_{\rm cal}$, of an estimated 20 per cent: 
$\sigma^2=\sigma_{\rm n}^2 + \sigma_{\rm cal}^2$.
Fourteen of the nineteen galaxies were detected at 850 $\mu$m, 
with 3$\sigma_{\rm n}$ upper limits of between 50 and 200 mJy obtained for the 
remaining five galaxies. These upper limits have been computed under
the assumption that the angular size of the galaxy is small compared
to the beam; if considerable extended emission is present, the 
total fluxes could be higher.

We have found extended emission for NGC 5653 and NGC 5936. For NGC 5653
a Gaussian fit yielded a Full Width Half Maximum (FWHM) of 17'',
whereas the shape of NGC 5936 is rather irregular with a maximum
diameter of about 40''. In the case of NGC 1614
and NGC 3110 there could be extended emission present, but the
noise level of our data is not low enough in order to definitely 
confirm this. For the other galaxies we determined  
the source sizes to have a FWHM of less than 8''. 
The flux at 850 $\mu$m was determined as the integrated flux
over the source extension. 

\begin{table}
\caption{FIR and submillimetre data}
\begin{center}
\begin{tabular}{lccccc}
\hline
\hline
(1) & (2) & (3) & (4) & (5) & (6)\\
Name & $\rm F_{12\mu m}$ & $\rm F_{25\mu m}$ & $\rm F_{60\mu m}$ &
 $\rm F_{100\mu m}$ & $\rm F_{850\mu m}$ $^{\rm (a)}$ \\
 & [Jy] &  [Jy] & [Jy] & [Jy] & [mJy] \\
\hline
NGC 1614  &   1.44 &   7.29 &  32.3 &  32.7 & 110 $\pm$  21$^{(3)}$ \\
NGC 3110  &   0.59 &   1.04 &  10.7 &  19.2 & 144 $\pm$  27$^{(2)}$ \\
NGC 4194  &   0.83 &   4.31 &  21.4 &  25.9 & 113 $\pm$  22$^{(1)}$ \\
NGC 4418  &   0.93 &   9.32 &  40.7 &  32.8 & 290 $\pm$  13$^{(1)}$ \\
NGC 5135  &   0.64 &   2.40 &  16.9 &  28.6 & 182 $\pm$  20$^{(1)}$ \\
NGC 5256  &   0.23 &   0.98 &   7.3 &  11.1 &  71 $\pm$   7$^{(1)}$ \\
NGC 5653  &   0.70 &   1.29 &  11.0 &  20.8 & 176 $\pm$  24$^{(1)}$ \\
NGC 5936  &   0.48 &   1.26 &   8.5 &  16.1 & 147 $\pm$  32$^{(1)}$ \\
NGC 6240  &   0.56 &   3.42 &  22.7 &  27.8 & 168 $\pm$  10$^{(3)}$ \\
Arp 193   &   0.26 &   1.36 &  15.4 &  25.2 & 104 $\pm$  16$^{(1)}$ \\
Arp 220   &   0.48 &   7.90 & 112.0 & 114.0 & 792 $\pm$  26$^{(1)}$ \\
Mrk 231   &   1.87 &   8.66 &  32.0 &  30.3 & 126 $\pm$  13$^{(1)}$ \\
Mrk 273   &   0.24 &   2.28 &  21.7 &  21.4 & 104 $\pm$  10$^{(1)}$ \\
Zw 049    &   0.08 &   0.77 &  20.8 &  29.4 & 180 $\pm$  16$^{(1)}$ \\
IC 2810   &   0.19 &   0.58 &   5.9 &  10.3 & $<$ 52$^{(2)}$ \\
NGC 1667  &   0.43 &   0.68 &   5.9 &  14.7 & $<$ 66$^{(3)}$ \\
NGC 2623  &   0.21 &   1.74 &  23.1 &  27.9 & $<$198$^{(3)}$ \\
IRAS 0519 &   0.73 &   3.44 &  13.7 &  11.4 & $<$110$^{(1,3)}$ \\
IRAS 1212 &   0.11 &   0.52 &   8.5 &  10.0 & $<$ 52$^{(1)}$ \\
\hline
\hline
\end{tabular}
\end{center}
$^{\rm (a)}$ The quoted flux error is estimated from the noise 
across the SCUBA maps, $\sigma_{\rm n}$. 
The upper limits are 3 $\sigma_{\rm n}$.

The chop throws used are (1) 60'', (2) 100'' and (3) 120''.
\end{table}

\section{Dust emission}

\begin{figure*}
\epsfig{file=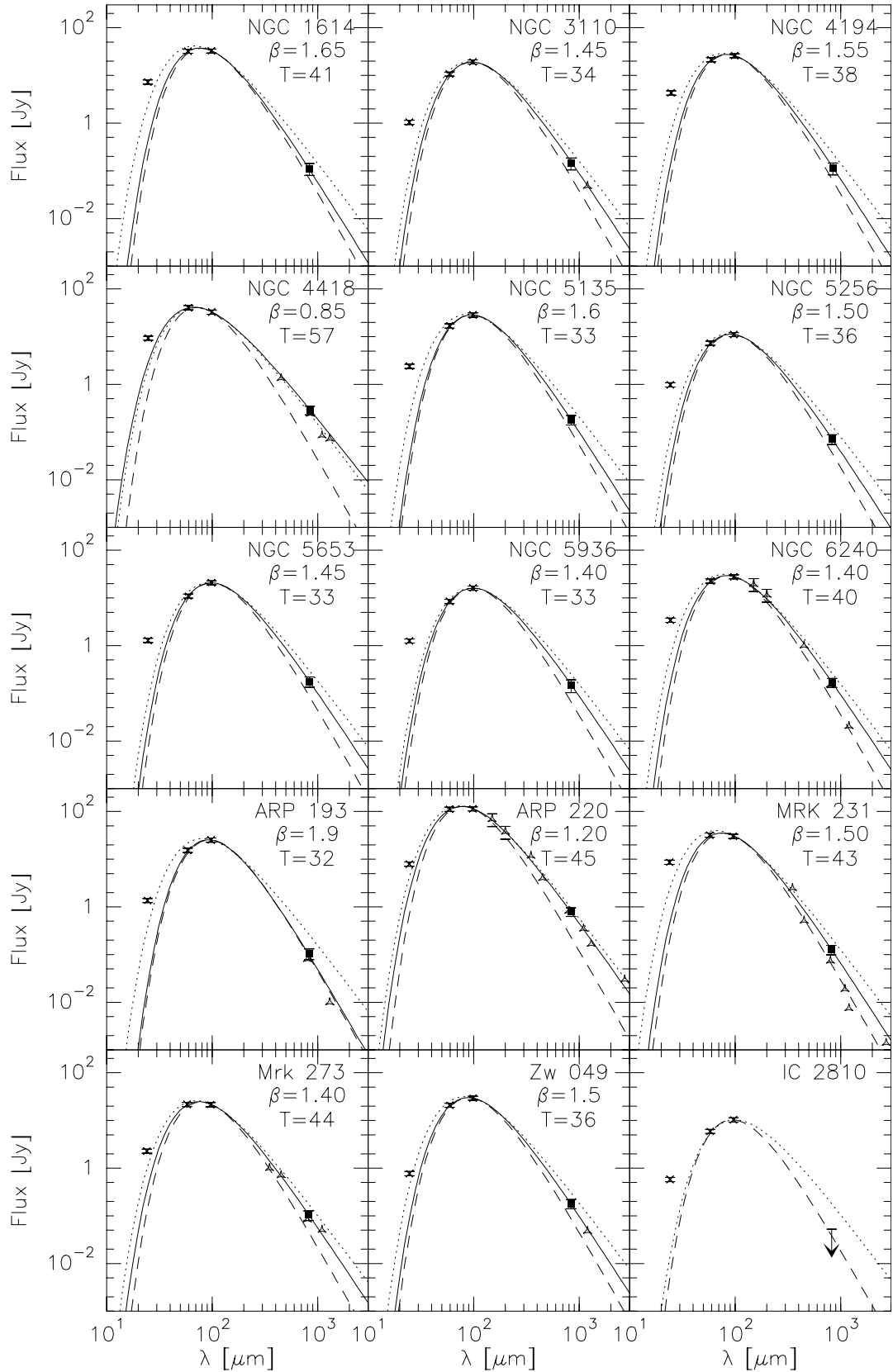,width=14.2cm,clip=,angle=0}
\caption{The dust emission spectrum for the galaxies.  
The filled squares are our 850 $\mu$m 
measurements, the crosses are IRAS data and the open triangle 
data taken from the literature (Braine \& Dumke 1998; 
Carico et al. 1992; Downes \& Solomon 1998;
Klaas et al. 1997;  Lisenfeld et al. 1996; Rigopoulou et al. 1996;
Roche \& Chandler 1993; Scoville et al. 1991).
The fits are  done taking
into account the data at 60, 100 and, in case of detection,
850 $\mu$m. The dashed line
is a fit with $\beta=2$, the dotted line for $\beta=1$ and in the
full line the fit is performed  also for $\beta$. The best-fit 
value for $\beta$ and for the dust temperature $T$
in this case are written in the graphs.}
\end{figure*}

\begin{figure*}
\epsfig{file=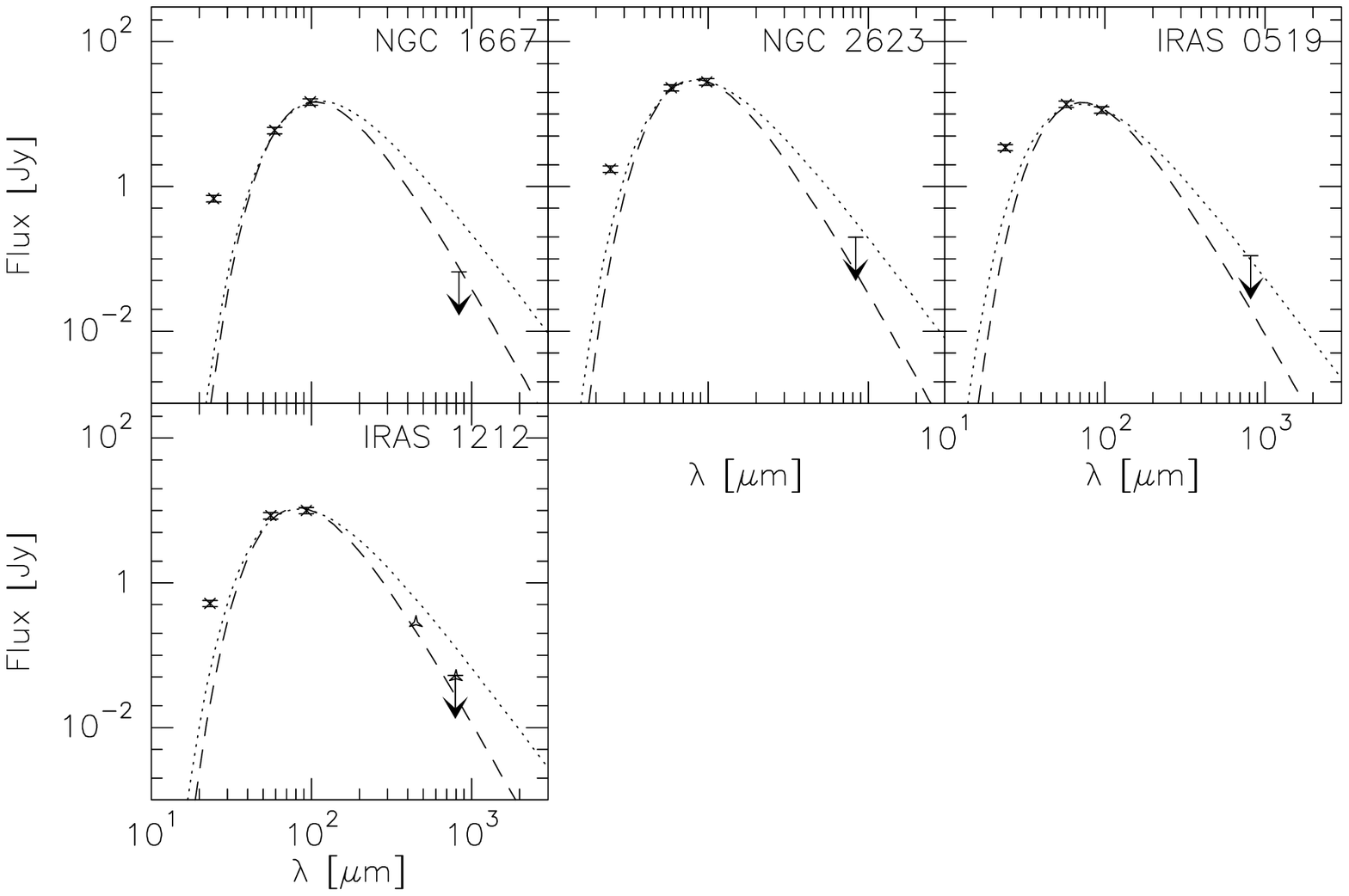,width=14.2cm, clip=,angle=0}
\centerline{ Figure 1 -- continued}
\end{figure*}

We have fitted simple
models to the observed 850 $\mu$m flux densities together with the 
IRAS flux densities at 100 $\mu$m, 60 $\mu$m, 
and,  in the case of the two-temperature model,
additionally at 25 $\mu$m.
The fits are made with only three or four data points 
and so it is not possible 
to constrain a model with more than this number of free parameters. 
Therefore, a number of assumptions are made about the dust. 
In the first model we assume that
the dust is at a single physical temperature, and that the emission 
is optically thin.
In the second model we fixed the dust emissivity law, and adopted a 
two-component optically thin dust model, while in the third, we explored 
an optically thick model.
In each case the data fits were made by varying the free model parameters 
and searching for the minimized $\chi^2$. 

\subsection{Optically thin, single-temperature emission}

We have used a single-temperature, optically thin dust model given by
\begin{equation}
S_\lambda = B_\lambda(T) \tau_\lambda \propto B_\lambda(T) k_\lambda
\end{equation}
to fit the 850, 100 and 60 $\mu$m flux densities, 
where $B_\lambda(T)$ is the Planck function, $\tau_\lambda$ 
is the dust opacity and $k_\lambda$ the 
dust absorption coefficient. 
At wavelengths long compared to the physical grain size ($\lambda \gg a$, 
with $a$ the grain size), the dust absorption coefficient
can be parameterized by 
\begin{equation}
k_\lambda\propto {\lambda}^{-\beta}.
\end{equation}
The wavelength dependence of $k_\lambda$ is not well known, and depends on the
physical composition of the dust. Theoretical arguments suggest values of 
$\beta$ between 1 and 2 
(see Tielens \& Allamandola 1987). Models of 
interstellar dust that are both astronomically realistic in composition,
and able to reproduce the observed dust extinction and
emission curves, predict $\beta \simeq 2$ (e.g. Draine \& Lee 1984, 
Ossenkopf \& Henning 1994, Kr\"ugel \& Siebenmorgen 1994). 
The value of $\beta$ becomes somewhat lower ($\beta \approx 1.5$,
D\'esert, Boulanger \& Puget 1990) if stochastic heating of 
small grains to high 
temperatures is taken into account.  Even smaller values of $\beta \simeq 0.4$ 
are predicted/observed in the atmospheres of Vega-type stars, 
where large dust grains that do not satisfy the condition of 
$\lambda \gg a$ even at submillimetre wavelengths can form
\cite{kruegel94}.

In Fig. 1 we show the data for the galaxies together with the best model fits. 
Three different curves are shown: the solid line depicts the 
dust emission model with the best-fit $\beta$ and $T$ while  the 
dotted and dashed lines show  the best-fit $T$ models 
with $\beta$ of 1 and 2 respectively. The dust-temperatures
derived for the case of variable $\beta$ lie between 30 and 60 K.
When only upper limits at 850 $\mu$m were measured, the
fits were made using a fixed $\beta$ and  the 60 and
100 $\mu$m flux densities only.
The results for $\beta$ and $T$ are not 
significantly different if an alternate   wavelength dependence
of $k_\lambda$  is used at short wavelengths ($k_\lambda \propto \lambda^{-1}$ 
for $\lambda < 250 \mu$m, Hildebrand 1983).

The dust emission from all galaxies is well-fit by the
single-temperature dust model, with $1\la \beta<2$.
The distribution of the values of the best-fit $\beta$  
as a function of $\lir/\lb$ are shown in Fig. 2, where the error bars on 
$\beta$ have been estimated by performing fits with the 
850 $\mu$m flux plus/minus $\sigma$.
The ratio $\lir/\lb$ is frequently used as an indicator for the 
star formation activity since $\lb$ and $\lir$ reflect the 
average star formation rates during the last $\simeq 3\,10^9$ and
$\simeq 10^8$ yr, respectively. 
Starburst galaxies have a much higher ratio ($\lir/\lb \simeq 1 -100$)
than normal galaxies ($\lir/\lb< 1$).
The interpretation of $\lir/\lb$ as an indicator of star formation has to be 
made with caution, as the ratio is also affected by the amount of
dust present  (Calzetti et al. 1995). No trend of $\beta$ with $\lir/\lb$
is seen.
%
The majority of galaxies in the sample have values of $\beta$ around
1.5 or higher, in agreement with the predictions
of astronomical models (see above). The only exceptions are  of Arp 220 and  NGC 4418  which
have a lower $\beta$, close to  $1$. 

These low values of $\beta$ could
be an indication of very different dust properties, for example, a higher
proportion of very small or very large grains. Although we cannot 
exclude this possibility, it is not clear why these two galaxies, 
that at face value do not differ substantially from the other members 
of the sample, should possess dust grains with different properties.
A more plausible explanation is that some of the simple assumptions on which 
our fits are based are not correct. In the following subsection we 
explore the validity of these assumptions.

\begin{figure}
\epsfig{file=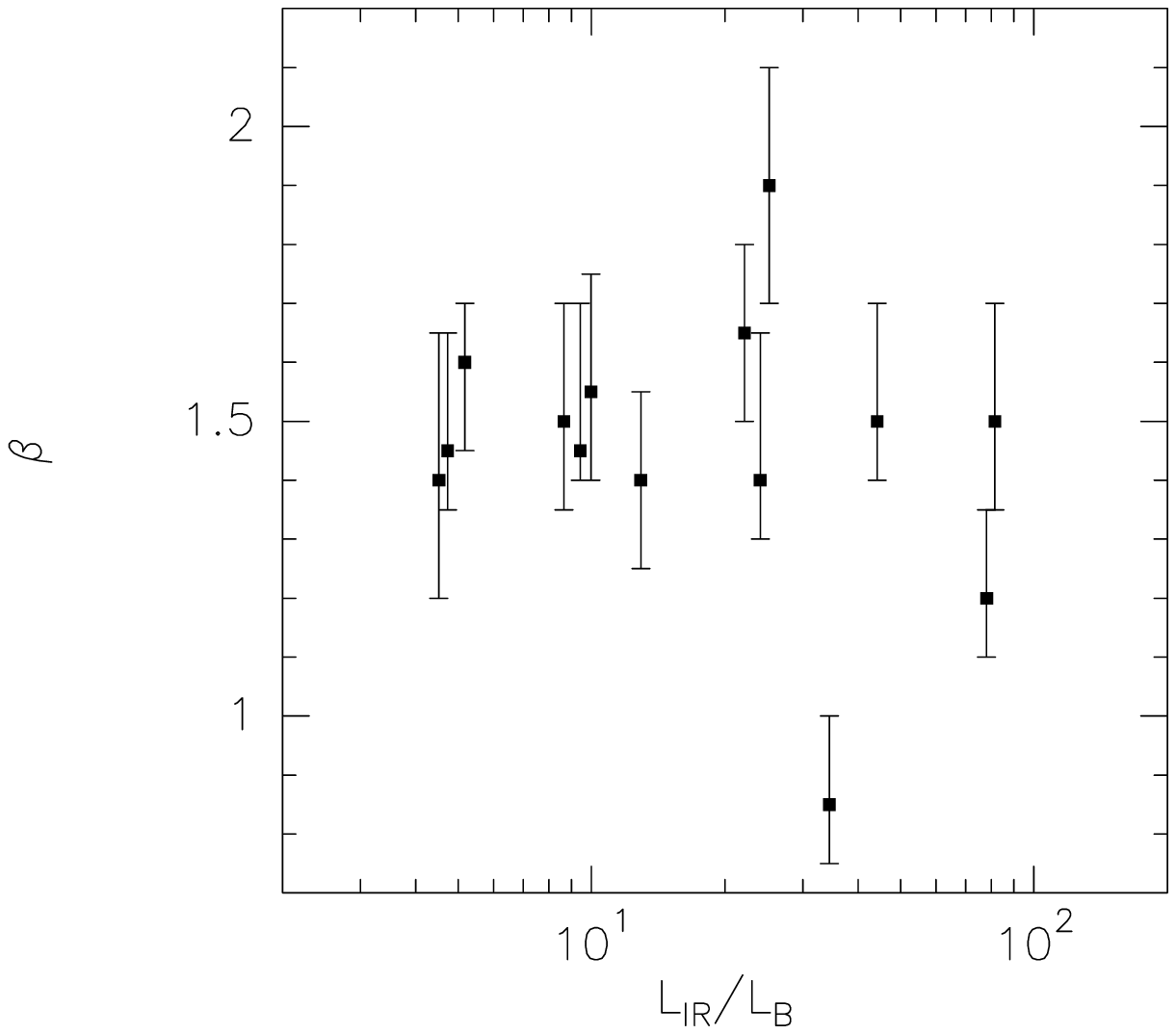,width=8.5cm, clip=,angle=0}
\caption{The exponent of the wavelength dependence of the
extinction coefficient, $\beta$, is shown as a function of
$\lir/\lb$.}
\end{figure}

\subsection{Optically thin, multi-temperature emission}

The modeling of the previous section was based on the assumption that all 
the dust was at a single dust temperature. This is quite simplistic 
as in principle the dust will be at a range of temperatures that  
depends on the local radiation field. 
Observations of spiral galaxies
have shown that indeed at least two components, a warm and a cold one
($T\la 20$K), are  necessary to 
explain the dust emission spectrum from  FIR to millimetre 
wavelengths (e.g. Gu\'elin et al. 1993,
Chini et al. 1995, Alton et al. 1998).
As a first approximation to a more 
realistic model, we have applied a two-temperature fit to the data,
with $\beta=2$.

In Fig. 3 we show the results of a two-temperature model fit to the  
25, 60, 100 and 850 $\mu$m fluxes for a selection of the galaxies in 
the sample: Arp~220 and NGC 4418 have been taken as examples of galaxies 
with low values of best-fit $\beta$, while NGC 6240 and Arp 193 are 
examples with higher values of $\beta$ (1.4 and 1.9,
respectively). The data points are reasonably well-fit for these as 
well as other  galaxies not shown. 
The discrepancy between the model and some of the data points, e.g. with the 
ISO fluxes at 150 and 200 $\mu$m of Arp 220
\cite{klaas}, could easily be improved
using more temperature components.

The high 850/60 and 850/100 $\mu$m flux ratios observed in 
Arp 220 and NGC 4418 necessitate the presence of
large amounts of cold dust. The temperature of the second dust component 
needed to model the emission from these two galaxies is
$T_{\rm c} \simeq 21 - 22$, rather colder than that typically 
found for active galaxies ($T=33$ K, Chini et al. 1995). 
In a more realistic multi-temperature model using more than 2 temperature
components, the temperature of the coldest 
dust component would be even lower.
This is surprising as, a priori, it is unlikely that Arp 220 and
NGC 4418 have colder dust than other galaxies in the sample, 
particularly given that they show among the highest values of
$\lir/\lco$\footnote{
The ratio $\lir/\lco$ can be interpreted 
as a rough measure of the dust temperature using the following argument:
$\lco$ is proportional to the brightness tempertature, $T_b$, of the
molecular gas in the galaxy.  If we assume that $T_b = T$, 
the temperature of the dust, and take $\lco$ as a measure of 
both the gas and dust mass (assuming
a fixed gas-to-dust ratio), we obtain after reordering
equation 7 and 
integrating over wavelengths, $\lir/\lco \propto
\int B_\lambda(T) k_\lambda {\rm d}\lambda /T \propto T^5$, 
for optically thin dust emission with
$\beta=2$. In contrast, if the dust is  optically thick, then this relation
changes to $\lir/\lco \propto
\int B_\lambda(T){\rm d}\lambda /T \propto T^3$. In both cases,
a high $\lir/\lco$ is  indicative of a high dust temperature.}.

\begin{figure*}
\epsfig{file=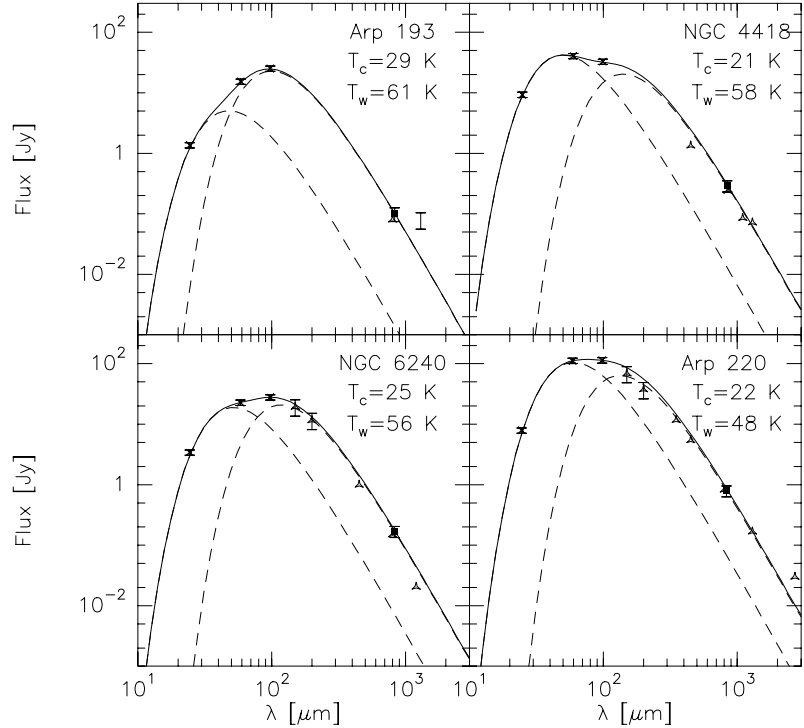,width=10.5cm, clip=,angle=0}
\caption{The dust emission  fitted with a two-temperature model
for $\beta=2$.
The temperature of the cold ($T_{\rm c}$) and of the warm ($T_{\rm w}$)
dust component are indicated.}
\end{figure*}

\subsection{Optically thick dust emission}

As discussed in the introduction, there is mounting evidence to suggest that
the dust opacity in LIRGs is high. To investigate the effects of dust 
opacity, the data at 60, 100 and 850 $\mu$m were fit using a dust 
emission model with a finite optical depth given by:  
\begin{equation}
S_\lambda=B_{\lambda}(T)\left(1-{\rm e}^{-\tau_\lambda}\right),
\end{equation}
where 
\begin{equation}
\tau_\lambda=\tau_{100\mu {\rm m}}(100 \mu {\rm m}/\lambda)^{\beta}. 
\end{equation}
%
\begin{figure*}
\epsfig{file=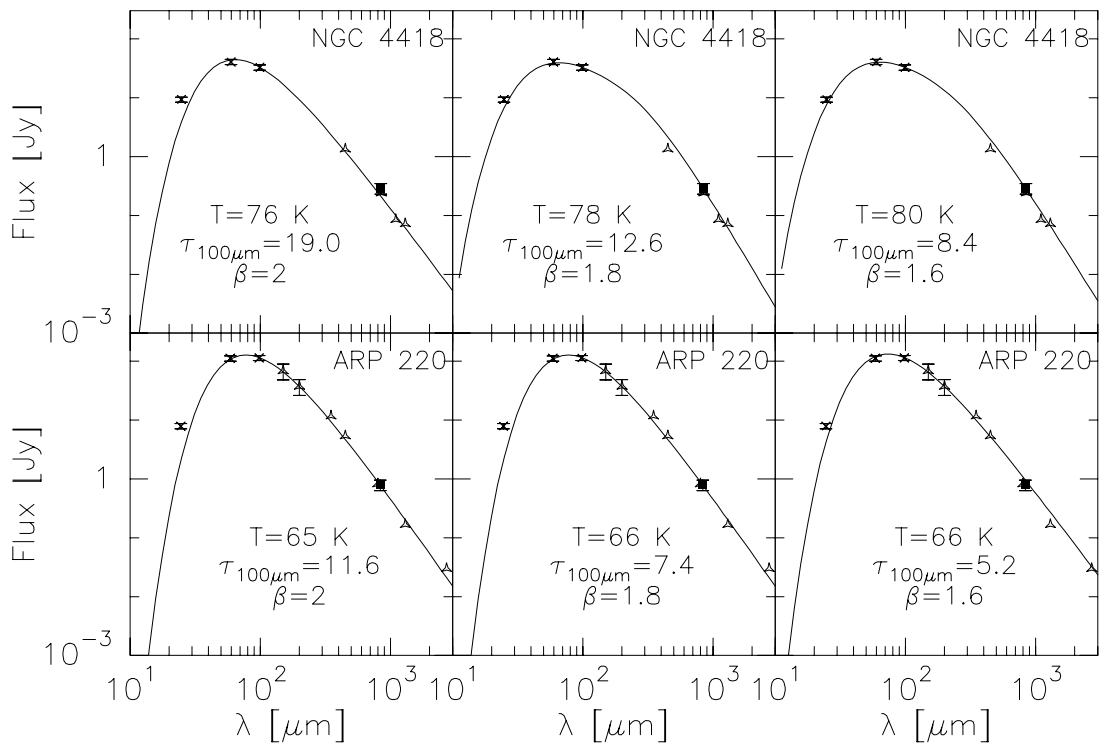,width=14.5cm, clip=,angle=0}
\caption{The dust emission, taking into account a finite
dust opacity, according to eqs. 3 and 4, is shown for Arp 220 and NGC 6240 for
different values of $\beta$. The best-fit value for the dust
opacity at 100 $\mu$m, $\tau_{100\mu {\rm m}}$, and the corresponding dust
temperature $T$ are indicated.}
\end{figure*}
%
The model fits for the two galaxies Arp 220 and NGC 4418 are shown in 
Fig. 4, where the dust temperature and opacity have been optimized for 
fixed values of $\beta$ (2, 1.8, 1.6).
For $\beta = 2$, the derived dust opacities are high and in Arp~220 
the fit is only marginally consistent with the observed 150 and 200 $\mu$m
fluxes. The fit improves with decreasing $\beta$, with the derived
opacities becoming consistent with independent estimates of other authors.

We conclude that the high 850/60 and 850/100 $\mu$m flux ratios 
of  Arp 220 and NGC 4418 relative to the rest of the sample is consistent with
a high dust opacity. 
 
The models discussed are very simplistic, and the 
true physical situation is most likely a combination of both a 
multiple-component dust temperature and a high dust opacity: a high dust 
opacity would result in the dust being shielded from the heating radiation,
resulting in cooler dust components in regions further from the radiation
source.

\section{Dust and gas masses}

The second objective of the project was to compare the molecular gas masses of
the sample derived using the two standard methods (a) from CO measurements
(b) from dust measurements.
\subsection{Molecular gas mass derived from the CO emission}
CO luminosities were calculated using CO(1-0) measurements taken 
from the literature. From these, we derived molecular gas masses 
using:
\begin{equation}
\mco=4.78 \lco  [\msol]
\end{equation}
\cite{sanders91}, with $\lco$, the CO luminosity in K km s$^{-1}$
pc$^2$, given by 
\begin{equation}
\lco=26.55 D^2 I_{\rm CO} \Theta_{\rm B}^2   [{\rm K\, km \,s^{-1} pc^2}]
\end{equation}
where $I_{\rm CO}$ is the integrated main beam temperature in K, $D$ is the 
source distance in Mpc, and $\Theta_{\rm B}^2$ the half power beam width 
(HPBW) of the beam in arcsec$^2$. The quantity $I_{\rm CO}$ has been 
taken from Sanders et al. \shortcite{sanders91}\footnote{
We have transformed the $T_{\rm R}^*$ temperature scale used in Sanders
et al. (1991) 
to $T_{\rm mb}$ by multiplication with $\eta_{\rm moon}/\eta_{\rm B}=1.19$ 
\cite{sanders91}} 
for the majority of galaxies, however, 
in the case of Arp 220, Arp 193, Mrk 231, Mrk 273 and NGC 6240 
data from Solomon et al. \shortcite{solomon97}
(taken with IRAM 30m telescope) were 
used instead. The fluxes measured by Solomon et al. \shortcite{solomon97} 
for 
three of these galaxies (Arp 220, Mrk 231 and Arp 193) are a factor of 
two higher than those measured by Sanders et. al \shortcite{sanders91}. 
The origin of this 
difference is unclear. We have adopted the values measured by
Solomon et al. \shortcite{solomon97} because of the 
higher signal-to-noise and better 
overall quality of their observed spectra. 
The values for $\lco$ and $\mco$ are tabulated in columns 4 and 7 of 
Table 2 respectively. The error in $\mco$ is determined by the error 
in $I_{\rm CO}$, which we have estimated by comparing the data used here with 
other published CO data \cite{young95}, where we find 
CO(1-0) fluxes for 11 galaxies of our sample. On average, there is 
a difference of about 25 per cent between the two data sets which we  
adopt as the uncertainty in $I_{\rm CO}$.

\begin{table*}
\caption{Luminosities and gas masses}
\begin{center}
\begin{tabular}{lccccccccc}
\hline
\hline
(1) & (2) & (3) & (4) & (5) & (6) & (7) & (8) & (9) & (10) \\
Name & Distance & $\lir$ & $\lco$ & $\lb$ & $\mbetabest$ & $\mco$ 
& $\frac{\lir}{\lco}$ & $\frac{\lir}{\lb}$ & $\frac{\mco}{\mbetabest}$ \\
  & [Mpc] & [$10^{11}\lsol$] & [$10^8\, L'$ ] &[$10^{10} \lsol$] &
[$10^9 \msol$] & [$10^9 \msol$] & & &  \\
\hline
NGC 1614  &   62.3 &   3.8 &   31.9 &    1.7 &   2.9 &  15.3 &  117.8 &   22.2 &   5.2 \\
NGC 3110  &   69.3 &   1.6 &   59.7 &    1.7 &   5.9 &  28.6 &   27.1 &    9.4 &   4.8 \\
NGC 4194  &   41.4 &   1.1 &    6.7 &    1.1 &   1.4 &   3.2 &  163.6 &   10.0 &   2.3 \\
NGC 4418  &   34.2 &   1.3 &    5.3 &    0.4 &   1.6 &   2.5 &  247.2 &   34.4 &   1.6 \\
NGC 5135  &   56.9 &   1.7 &   34.4 &    3.3 &   5.2 &  16.4 &   49.2 &    5.2 &   3.2 \\
NGC 5256  &  115.9 &   2.9 &   63.0 &    3.3 &   7.6 &  30.1 &   45.6 &    8.7 &   3.9 \\
NGC 5653  &   54.4 &   1.1 &   15.0 &    2.3 &   4.5 &   7.2 &   72.1 &    4.7 &   1.6 \\
NGC 5936  &   59.0 &   1.0 &   19.0 &    2.2 &   4.4 &   9.1 &   52.4 &    4.5 &   2.0 \\
NGC 6240  &  100.9 &   6.4 &   90.4 &    4.9 &  12.1 &  43.2 &   70.4 &   12.9 &   3.6 \\
Arp 193   &   97.8 &   4.0 &   44.3 &    1.6 &   9.1 &  21.2 &   91.3 &   25.2 &   2.3 \\
Arp 220   &   79.2 &  15.8 &   88.0 &    2.0 &  30.0 &  42.0 &  179.9 &   78.2 &   1.4 \\
Mrk 231   &  173.9 &  31.0 &   85.6 &    7.0 &  24.2 &  40.9 &  361.8 &   44.2 &   1.7 \\
Mrk 273   &  153.8 &  12.2 &   57.8 &    5.1 &  15.3 &  27.6 &  211.8 &   24.1 &   1.8 \\
Zw 049    &   52.3 &   1.3 &    8.9 &    0.2 &   3.8 &   4.3 &  149.9 &   81.5 &   1.1 \\
IC 2810   &  140.5 &   3.2 &   54.8 &    3.4 & $<$ 10.0 &  26.2&   57.6 &    9.3 & $>$  2.6 \\
NGC 1667  &   58.9 &   0.8 &   12.3 &    3.4 & $<$  2.5 &   5.9&   62.3 &    2.3 & $>$  2.3 \\
NGC 2623  &   76.1 &   3.2 &   26.6 &    2.8 & $<$  9.4 &  12.7&  121.1 &   11.7 & $>$  1.4 \\
IRAS 0519 &  167.5 &  11.7 &   69.8 &     -- & $<$ 21.2 &  33.4&  167.4 &    -- & $>$  1.6 \\
IRAS 1212 &  293.3 &  16.7 &  123.5 &     -- & $<$ 36.7 &  59.0&  135.1 &    -- & $>$  1.6 \\
\hline
\hline
\end{tabular}
\end{center}

\flushleft{
(2) Distance taken from Sanders et al. \shortcite{sanders91}, 
based on $H_0=75$  km s$^{-1}$
Mpc$^{-1}$ and taking into account the Virgocentric flow.

(3) $\lir=L(8-1000\mu{\rm m})=5.6\,10^5 D_{\rm Mpc}^2(13.48 F_{12}+ 
5.16 F_{25}+
2.58 F_{60}+F_{100})$ where the flux is in Jy \cite{sanders96}.

(4) $L'$=K km s$^{-1}$ pc$^2$; $\lco$ is calculated according to eq. (6).

(5) Calculated from $B_{\rm T}^0$, taken from de Vaucouleur et al. 
\shortcite{RCS},
except for Zw 049 and NGC 3110 which have been taken from
Lehnert \& Heckman \shortcite{lehnert} and 
Lutz \shortcite{lutz}, respectively.

(6) Molecular gas mass estimated from the CO emission, calculated 
according to eq. (5).

(7) Molecular gas mass estimated from the dust emission, calculated 
according to eq. (7).
}

\end{table*}

\subsection{Molecular gas mass derived from the dust emission}

The molecular gas mass can be calculated from the 850
$\mu$m emission:

\begin{equation}
\mbetabest =\frac{D^2 S_{850\mu{\rm m}}}
{B_{850\mu{\rm m}}(T) k_{850 \mu{\rm m}}}
\end{equation}
where optically thin dust emission at 850 $\mu$m is assumed (this
is correct even for the high dust opacities estimated in the 
previous section).

We have 
adopted temperatures derived in the best-fit $\beta$ models. 
We have used the value for $k_{850 \mu{\rm m}}$ given by Hildebrand 
\shortcite{hildebrand} 
which, for $\beta=2$ and for a gas-to-dust mass ratio of 150, is 
$k_{850 \mu{\rm m}}=0.0058\, \rm cm^2 g^{-1}$. For comparison 
with other authors, 
this value can be extrapolated to 1.3 mm, where we obtain (for $\beta=2$) 
$k_{1.3{\rm mm}}=0.0025 \,\rm cm^2 g^{-1}$. 
Based on extensive simulations of dust emission and absorption 
Kr\"ugel, Steppe \& Chini \shortcite{kruegel90} estimate 
$k_{1.3{\rm mm}}=0.003 \,\rm cm^2 g^{-1}$.
Mezger, Wink \& Zylka \shortcite{mezger} suggest a value of 
$k_{1.3{\rm mm}}=0.0013 \, b\,z \rm \,cm^2 g^{-1}$, where $b$ is a measure of 
the dust environment and $z$ is the metallicity.
Adopting a value of $b=1.9$ for moderately dense gas
yields (for $z=1$) a value of $k_{1.3{\rm mm}}=0.0025 \,\rm cm^2 g^{-1}$, 
which matches the 
value adopted in this paper. In contrast, the ``standard'' 
dust model of Draine \& Lee \shortcite{draine} gives a somewhat lower
value of $k_{1.3{\rm mm}}=0.0015 \,\rm cm^2 g^{-1}$, which has been found to 
agree with observational comparisons between dust extinction at 
IR wavelength and dust emission at FIR/mm wavelength in dense Galactic 
molecular cloud cores (Kramer et al. 1998, Lehtinen et al. 1998).

The most significant uncertainty in this method,
apart from $k_{850 \mu{\rm m}}$, is the dust temperature.
The error in the dust temperature can be estimated by comparing
the temperatures derived in the different models
(optically thin/thick, single temperature
or various temperature components) used here, which differ by
about 50 per cent.
At submillimetre wavelengths the Planck function depends linearly on 
the dust temperature, so this uncertainty in the dust temperature
produces an error in $\mbetabest$ 
of 50 per cent also. 
Adding to this quadratically the measurement error of the  850 $\mu$m
flux densities (on average 25 per cent), we estimate an error of
55 per cent for $\mbetabest$. 

\subsection{Comparison between the two masses}

\begin{figure}
\epsfig{file=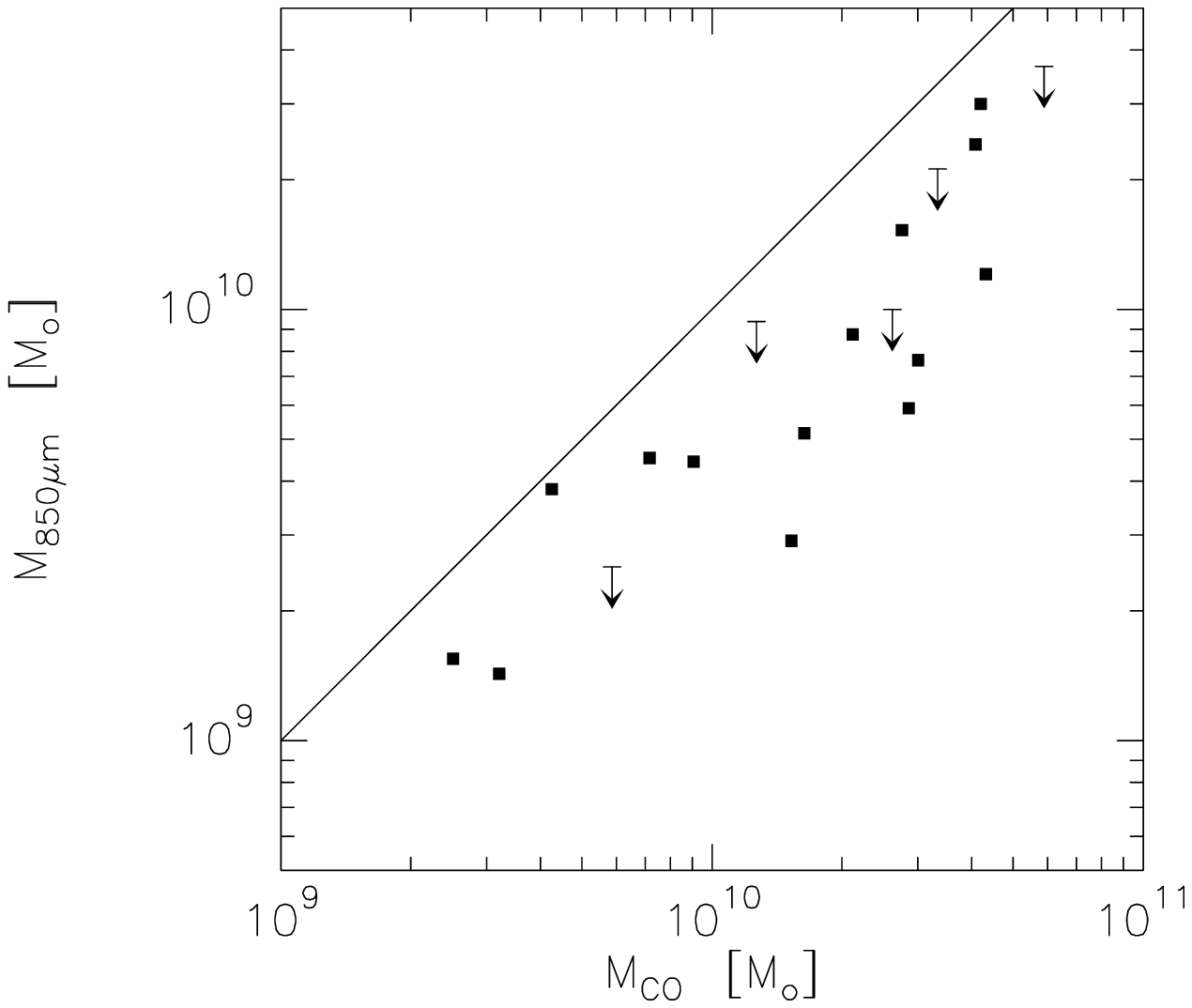,width=8.5cm, clip=,angle=0}
\caption{The molecular gas mass derived from the CO-luminosity, $\mco$,
and the molecular gas mass derived from the dust emission, $\mbetabest$.
The diagonal line shows the relation $\mco = \mbetabest$.}
\end{figure}

In Fig. 5 we show the comparison between the molecular gas mass derived
from the CO emission, $\mco$, and the molecular gas mass derived
from the 850 $\mu$m continuum, $\mbetabest$. 
$\mco$ is systematically higher than $\mbetabest$.
A good correlation 
(correlation coefficient $r=0.82$) exists 
between the two quantities, spanning nearly 2 orders of magnitude.

In Fig. 6 we have plotted the ratio between the molecular gas mass derived 
using the two alternative methods as a function of $\lir$
and as a function of $\lir/\lb$   
to see whether there is any trend with luminosity 
or the star formation activity.
The error of $\mco/\mbetabest$
is the root of the sum of the squares of the errors 
in $\mco$ and $\mbetabest$, and so approximately 60 per cent.
The average ratio, $\mco/\mbetabest$ is $2.6 \pm 1.3$, ignoring upper limits.
There is no trend visible for $\mco/\mbetabest$ to vary with $\lir$.
There might be a weak anticorrelation in the LIRG sample
between  $\mco/\mbetabest$ and $\lir/\lb$, 
but the statistical significance is low
and a larger sample size will be needed in order to confirm or
refute this anticorrelation.

\begin{figure*}
\centerline{\epsfig{file=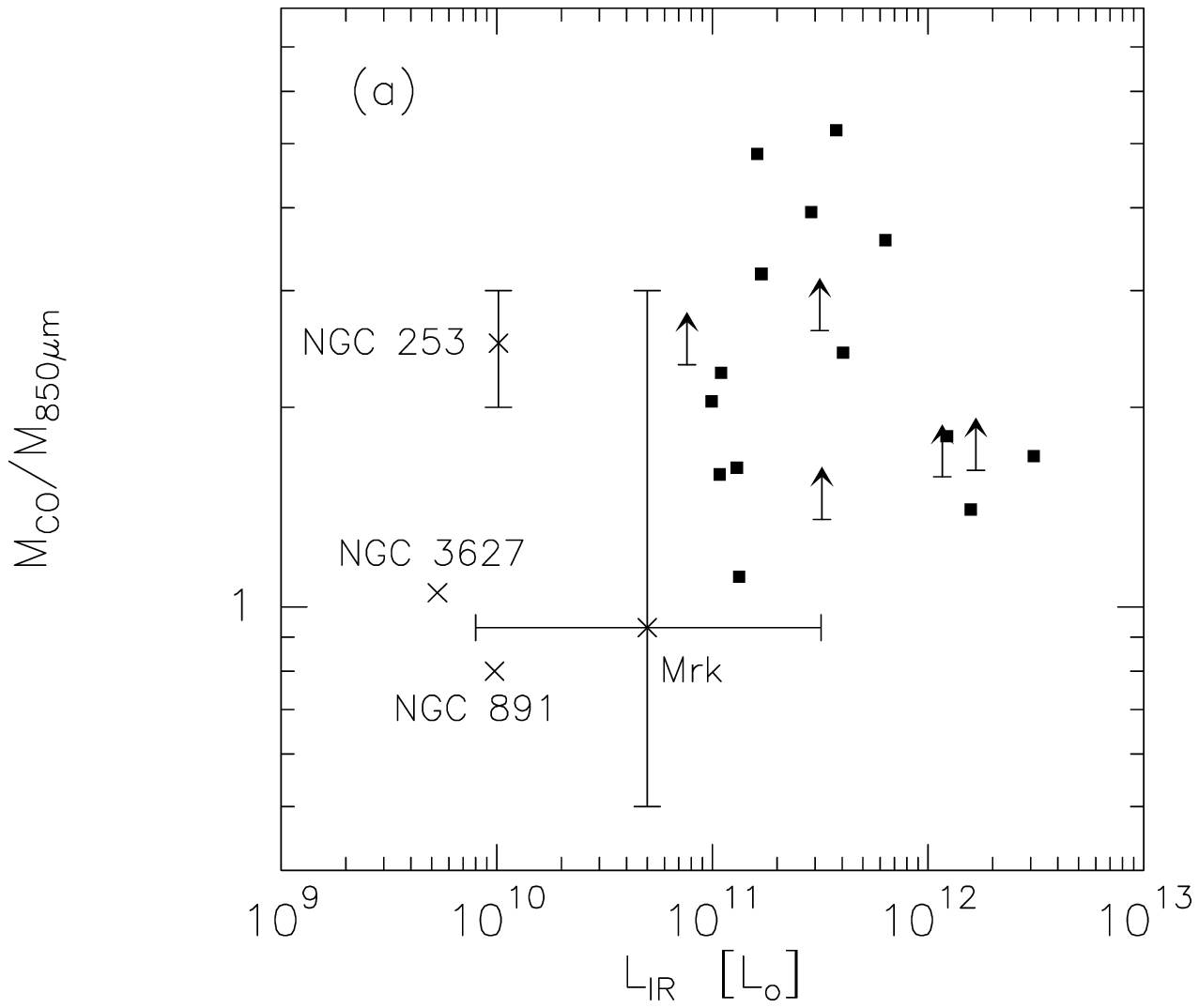,width=8.5cm, clip=,angle=0} \quad
\epsfig{file=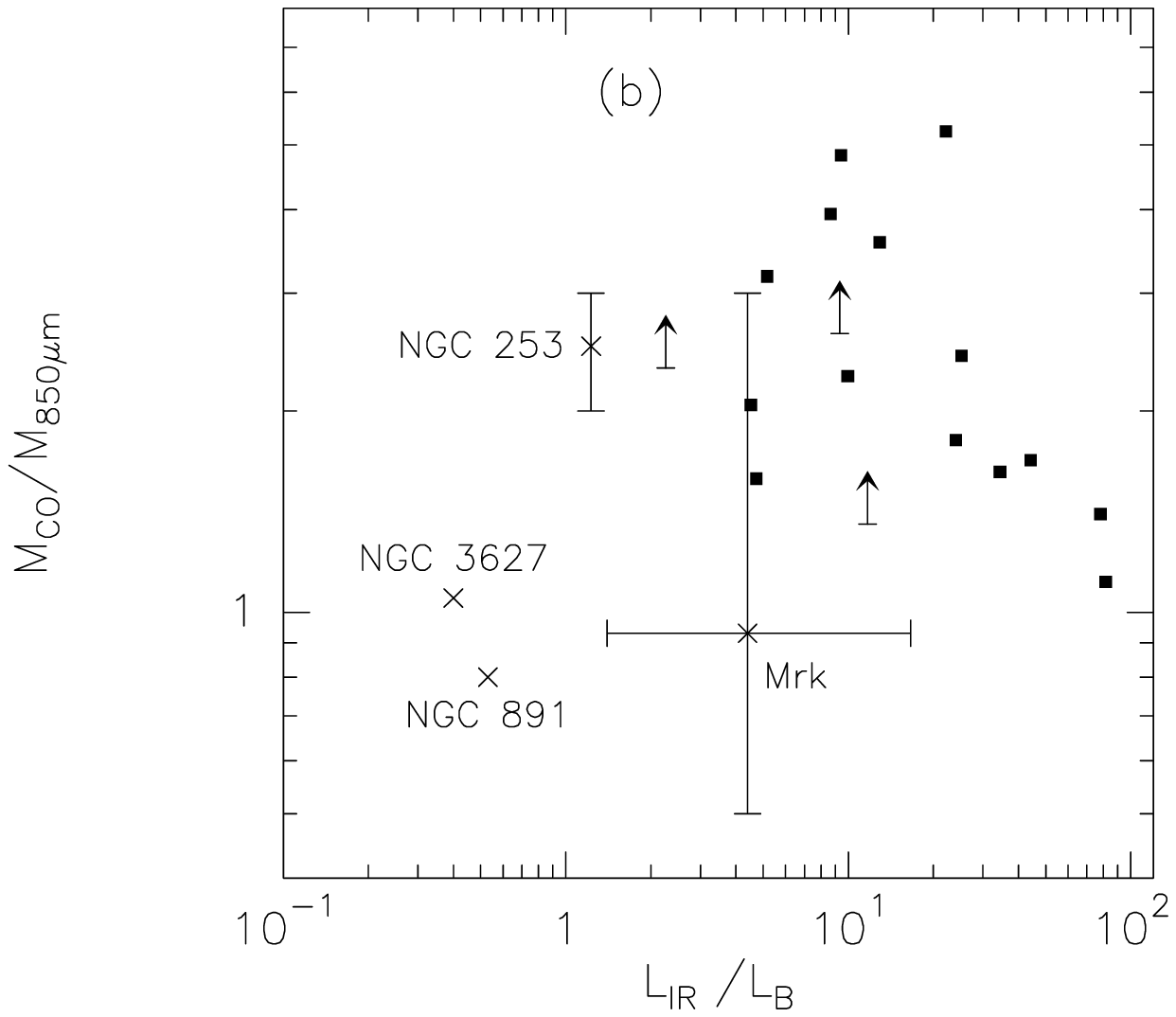,width=8.5cm, clip=,angle=0}}
\caption{The ratio between the molecular gas mass derived from the
CO emission and the molecular gas mass derived from the dust 
emission as a function
of $\lir$ (a) and of $\lir/\lb$ (b) for 
the LIRG sample (filled squares and arrows). 
For comparison, data from the literature (crosses) for spiral galaxies 
(NGC 3627, Sievers et al. 1994; NGC 891, 
Gu\'elin et al. 1993), 
Markarian galaxies (``Mrk'',the error bar gives the range for 
the values in the sample; Chini, Kr\"ugel \& Steppe 1992), 
and the starburst galaxy
NGC 253 (the error bars give the range of values found
in the center and in the disk; Mauersberger et al. 1996) 
are also included.}
\end{figure*}

Also plotted in Fig. 6 are data taken from the literature for 
other galaxies, including spiral galaxies (NGC 3637, NGC 891) and 
actively star forming and starburst galaxies (Markarian galaxies, NGC 253). 
The $\mco/\mbetabest$ ratios for these comparison galaxies
have been rederived using the same 
parameters as for the LIRG sample.  
The most significant differences between the LIRG sample  
and these other galaxies is seen in the
spiral galaxies NGC 891 (factor of 3.3) 
The starburst galaxy NGC 253 has about the same ratio as 
the average of the LIRGs, while the Markarian galaxies have a lower value,
possibly
due in part to the fact that $\mco$ has been 
derived from the CO(2-1) line emission assuming a ratio 
of CO(2-1)/CO(1-0) of 1. 

We have found for the LIRG sample that the molecular gas mass 
derived from the CO is higher than
that derived from the 850 $\mu$m. This confirms the conclusion 
of other authors, that the standard galactic CO-to-molecular mass conversion factor 
overestimates the molecular gas mass in LIRGs. 
The resultant ratio of $\mco/\mbetabest$ is slightly less than that
seen in other studies, the results of which suggest that the 
CO emission overestimates the molecular gas mass by a factor
of about 4-5 (Solomon et al. 1997, Downes \& Solomon 1998) or
more (factor 3-10, Shier et al. 1994, Braine \& Dumke 1998). If this
higher value is true, it implies that the molecular gas mass derived from 
the 850 $\mu$m flux is also an overestimate. 
This could be possibly  due to
a higher dust-to-gas mass ratio in these galaxies 
which could be caused by the higher metallicity produced
by intense star formation as seen in galactic nuclei and
bulges, for example in the inner part of M~51 \cite{guelin95}
and the centre of NGC 253 \cite{mauersberger}.

\section{Conclusions and Summary}

We have presented new data at 850 $\mu$m 
taken with SCUBA at the JCMT for 19 luminous
infrared galaxies. Fourteen galaxies were detected
at a level of 5$\sigma$ or better. 
Using the 850$\mu$m data as well as IRAS fluxes taken from the literature, 
we have shown that the submillimetre/far-infrared spectral energy distribution 
for the majority of the sources in the sample is well-described by an 
optically thin, single temperature dust model, with  
$\beta \simeq 1.4 - 2$. A  lower value of $\beta \simeq 1$ is required 
to fit the data for two of the galaxies, 
Arp 220 and NGC 4418. 

This difference can be attributed to one, or
a combination of:
(i) different intrinsic dust properties, (ii) the presence of a large
cold-dust component, (iii) high dust opacities.
We have discussed the various possibilities and conclude that the
most physical explanation is a high dust opacity. 

We have compared the molecular gas masses derived from the dust 
emission to the molecular  gas masses derived from the CO emission of our 
sample.  We find, on average, that $\mco$ is a factor of 2--3 higher than 
$\mbetabest$.  If the CO luminosity indeed overestimates the
gas mass by a factor of about 5-10 as indicated by other studies, then 
the dust emission must also overestimate the gas mass.
The most likely reason for this is a higher dust-to-gas
mass ratio in LIRGs.

\section*{Acknowledgments} 
We would like to thank the staff at the JACH
for carrying out the observations, Rob Ivison for
useful advices with respect to the data  analysis and   
and Loretta Dunne for allowing us 
to compare our results with observations made in the  
SCUBA Local Universe and Galaxy survey prior to publication.
We would also like to thank the referee for very helpful comments.

\end{document}